\documentclass[conference]{IEEEtran}
\IEEEoverridecommandlockouts

\usepackage{cite}
\usepackage{amsmath,amssymb,amsfonts}
\usepackage{algorithmic}
\usepackage{graphicx}
\usepackage{textcomp}
\usepackage{xcolor}
\usepackage{listings}

\usepackage{amsmath,amssymb,amsthm}
\usepackage{url}
\usepackage[ruled, vlined, linesnumbered]{algorithm2e}
\usepackage[english]{babel}
\usepackage{mathtools}
\usepackage{caption}
\newcommand{\blue}[1]{\textcolor{blue}{#1}}

\newcommand{\name}{\texttt{Mnemonic}}
\newcommand{\idx}{\texttt{DEBI}}
\newcommand{\idxroots}{\emph{roots}}
\newcommand{\matcher}{\emph{edgeMatcher}()}
\newcommand{\enumerator}{\emph{enumerator}()}

\newcommand{\netflow}{\emph{NetFlow}}
\newcommand{\lsbench}{\emph{LSBench}}
\newcommand{\lanl}{\emph{LANL}}

\def\BibTeX{{\rm B\kern-.05em{\sc i\kern-.025em b}\kern-.08em
		T\kern-.1667em\lower.7ex\hbox{E}\kern-.125emX}}

\begin{document}
\title{{\name}: A Parallel Subgraph Matching System for Streaming Graphs\\
\thanks{This work was supported in part by DARPA under agreement number N66001-18-C-4033 and National Science Foundation grants 1618706, 1717774, and 2127207. The views, opinions, and/or findings expressed in this material are those of the authors and should not be interpreted as representing the official views or policies of the Department of Defense, National Science Foundation, or the U.S. Government.}
}
\author{\IEEEauthorblockN{Bibek Bhattarai,    Howie Huang}
	\IEEEauthorblockA{\textit{George Washington University} \\
		\{bhattarai\_b, howie\}@gwu.edu}
}

\maketitle

\begin{abstract}
	Finding patterns in large highly connected datasets is critical for value discovery in business development and scientific research. This work focuses on the problem of subgraph matching on streaming graphs, which provides utility in a myriad of real-world applications ranging from social network analysis to cybersecurity. 
	Each application poses a different set of control parameters, including the restrictions for a match, type of data stream, and search granularity.  The problem-driven design of existing subgraph matching systems makes them challenging to apply for  different problem domains.
	This paper presents {\name}, a programmable system that provides a high-level API and democratizes the development of a wide variety of subgraph matching solutions.
	Importantly, {\name} also delivers key data management capabilities and optimizations to support real-time processing on long-running, high-velocity multi-relational graph streams. 
	The experiments demonstrate the versatility of {\name}, as it outperforms several state-of-the-art systems by up to two orders of magnitude. 
\end{abstract}

\begin{IEEEkeywords}
	subgraph, graph pattern, matching, isomorphism, streaming 
\end{IEEEkeywords}

\vspace{-2mm}
\section{Introduction}
\label{sec:introduction}
Subgraph matching is the task of finding matches of a specific pattern in a \textit{data graph}. Those patterns are referred to as \textit{query graphs} and each of their matches in the data graph are known as \textit{embeddings}.  Subgraph matching has several applications such as social media analysis~\cite{erling2015ldbc}, fraud detection~\cite{fraud_ring_neo4j}, and cyber-attack detection~\cite{data_breach_rep2018}. Due to constant data generation in these systems, promptly finding the newly formed embeddings is important in order to get new insights. 

However, there are still some major challenges to build a  subgraph matching system for streaming graphs. Most of the existing works follow filtering-enumeration technique~\cite{sun2019inmemory}, where per-query metadata, generally referred as an \textit{index}, is used to store the candidate matches of the query graph and provide better access locality for backtracking processes~\cite{Han:2013:TIT:2463676.2465300, He:2008:GQL:1376616.1376660, Ren:2015:EVR:2735479.2735493, bi2016efficient,bhattarai2019ceci,han2019efficient,sun2020rapidmatch}. Later, all the embeddings are listed and verified by traversing the index. However, these works are not suitable for streaming graph as updating index for single edge insertion or deletion can take up to $\mathcal{O}{(|V|)}$, where $V$ is the number of nodes on the data graph. Recomputing the index every time the graph gets updated incurs huge redundant computation. Recent works~\cite{kim2018turboflux,choudhury2015selectivity} present data-graph centric index structures to improve the updating and incremental subgraph matching.  However, there is still a large gap to be filled in order to support a real-time subgraph matching (1) The multiple edges between two endpoints(e.g., NetFlow events at different time) are treated as one entry in their index and thus loses the temporal context of the system behavior. (2) While graphs are streamed, the index is maintained entirely in the memory. 
Since index size is much bigger than graph itself, it is not suited for long running streams with large data volume. (3) The edge insertions and deletions are processed in strictly sequential manners. While there is room for parallelization even for 1-edge insertion or deletion, it fails to sufficiently utilize the computing power of multi-core machines.

In addition, there is a lack of general framework for subgraph matching that can be tailored to the problem in hand. Most datasets, in addition to edge and node type contain multiple attributes per node/edge and the combination of these features are used to match a data graph edge to a query edge. Similarly, the definition of \textit{match} itself varies depending upon the data and problem at hand. Some of the commonly used matching techniques include \textbf{subgraph isomorphism}~\cite{Han:2013:TIT:2463676.2465300,kim2016dualsim,bhattarai2019ceci,han2019efficient},  \textbf{homomorphism}~\cite{kim2015taming,kim2018turboflux,ammar2018distributed,kankanamge2017graphflow}, and \textbf{simulation}~\cite{ma2011capturing,fan2013incremental}. Most of the existing works hard-code these specifications in their system.
Without expertise in graph analysis, it is very difficult for users to choose appropriate subgraph matching solutions or make necessary changes. 
We present {\name}, a programmable subgraph matching framework for streaming graphs. It is designed with following objectives: (1) It distinguishes the different instances of edges between the same endpoints, enabling a context-aware subgraph matching. (2)  It supports incremental computation on long running, high-velocity event streams using batch processing and disk support whenever necessary, and (3) It exposes a high-level API set to democratize the implementation of subgraph matching solution that fits user requirements.

In order to accomplish these goals, {\name} designs the index storage, henceforth known as \textbf{D}ata graph  \textbf{E}dge centric, \textbf{B}inary \textbf{I}ndex ({\idx}), which uses a bitmap to represent whether a given data graph edge matches the query graph edges. In order to support multi-graph, each edge in the graph is assigned a unique identifier, \emph{edgeId}, and the index along with other attributes for each data graph edge is accessed using \emph{edgeId}. The {\idx} entry for a given data-query edge pair can be read, written, and updated in constant time($\mathcal{O}(1)$). By reclaiming memory from deleted edges and their {\idx} entries and periodic resets, memory consumption increases at a much smaller rate as the stream progresses. To the best of our knowledge, {\name} is the first subgraph matching system that achieves a non-monotonic index size. For events requiring a bigger search context, {\name} backups the data edges and their corresponding matches in the disks.

In order to batch the updates, {\name} uses a snapshot generator with a set of user-controlled parameters, e.g., stream type, window size, and stride. Each snapshot includes the last instance of the data graph and the changes made since then. Based on the insertions/deletions, the {\idx} is updated and new embeddings are enumerated. For each inserted/deleted edge, this process has to explore graph top-down and bottom-up to find all the impacted matches. Since, there is huge amount of redundant graph traversals among the exploration for each edge, {\name} constructs a unified traversal frontier where each edge is traversed only once for a batch of insertion or deletion. Since the batch of edges are processed at once, redundant embeddings are generated, which are proactively eliminated using masking. 

Finally, to ease the process of implementing subgraph matching solutions, it exposes a set of high-level APIs. For the majority of subgraph matching variants, a user needs to implement two application-specific functions - {\matcher} function controls the content of {\idx} by defining the matching conditions between a data edge and a query edge based on the vertex and edge attributes, and {\enumerator} function accesses the {\idx} using APIs to construct the embeddings that meet the structural constraints of the query graph. Optimization routines such as finding match order and starting nodes and parallelization are done by {\name}. Experienced users can achieve more flexibility by modifying these optimizations routines as well. To the best of our knowledge, {\name} is the first programmable subgraph matching framework.


\begin{figure}[t]
	\begin{center}
		\vspace{-0.1in}
		\includegraphics[scale=0.33]{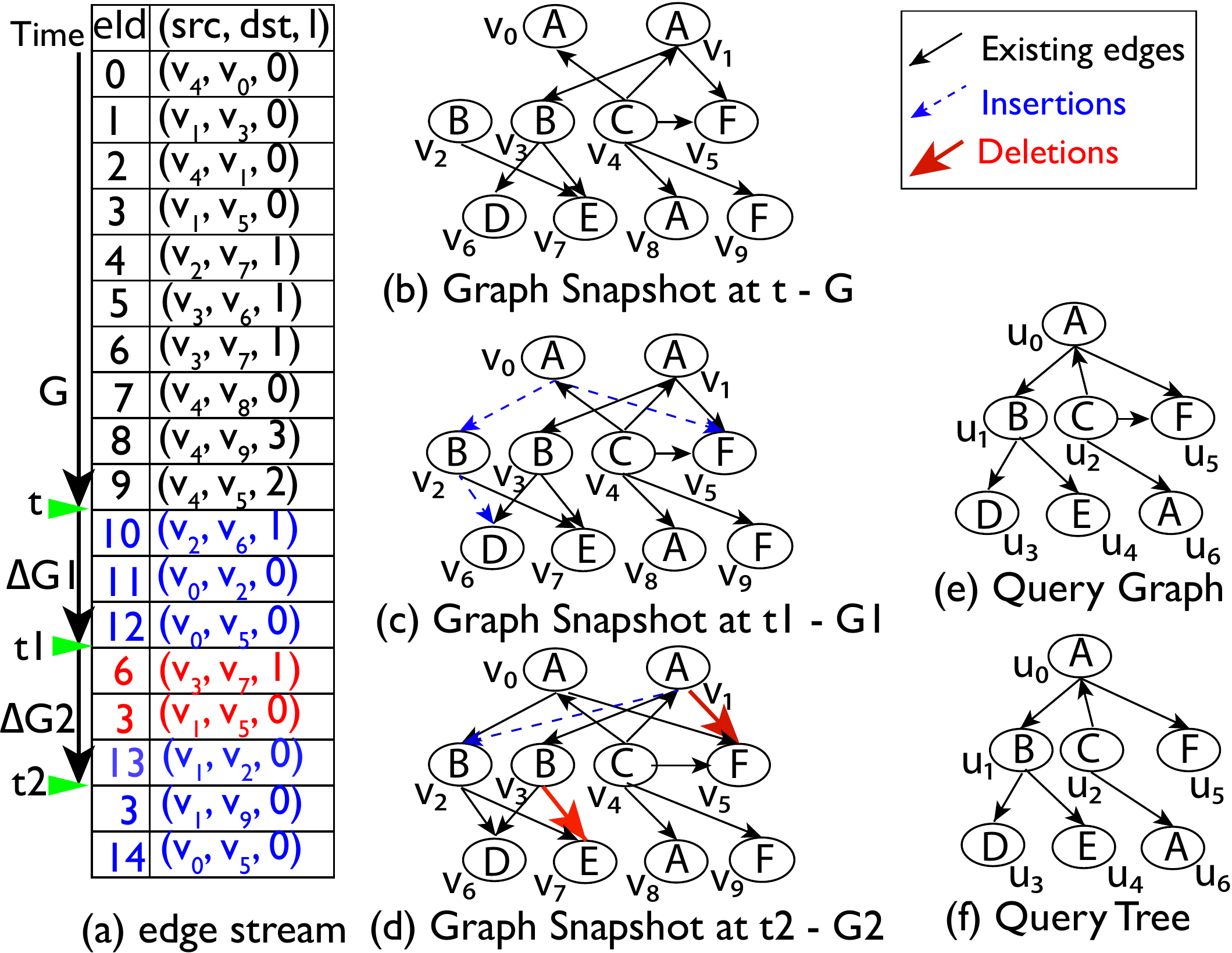}
		\caption{Example query and data graph pair.  The query graph has empty label on each edge, i.e., they match any label.}
		\label{fig:graphs}
	\end{center}
	\vspace{-0.2in}
\end{figure}





We have implemented several subgraph matching variants using {\name} with different embedding definitions and data characteristics.
A series of evaluations demonstrate that {\name} is able to outperform state-of-the-art systems on various problems, e.g., streaming graph isomorphism TurboFlux~\cite{kim2018turboflux} by $7.8\times$ and BigJoin~ \cite{ammar2018distributed} by a orders magnitude in larger queries, re-computation from scratch using static graph solution CECI~\cite{bhattarai2019ceci} by $42.1\times$, and time-constrained isomorphism~\cite{li2019time} by $1.8\times$. 
\vspace{-2mm}
\section{Background}
\label{sec:background}
\subsection{Preliminaries}

A graph is formally defined as $G = (V, E, L^v, L^e)$, where $V$ is the set of vertices, $E$ the set of edges, and $L^v$ and $L^e$  attribute functions that map a node and an edge to a set of attributes $\Xi^v$ and $\Xi^e$ respectively. 
Most commonly used edge attributes in subgraph matching are labels, which represent the types of nodes or relationships. 
Figure~\ref{fig:graphs}(b), (c) and, (d) show the different snapshots of an example data graph, i.e., the initial graph $G$ at time $t$ and two updated snapshots $G1 = G \oplus \Delta G1$ and $G2 = G1 \oplus \Delta G2$ at time $t1$ and $t2$ respectively. 
We use directed graphs in the examples and experiments, but our techniques work on undirected graphs as long as both directions are recorded. To facilitate the easy update and sequential accesses, {\name} uses the \textbf{adjacency list} format to store the graphs, where each vertex has a list that stores all its outgoing and incoming edges. To make sure the edges in multigraph are differentiated, each edge is assigned an id. The vertex and edge attributes are stored in another data structure indexed by their id.
 
A subgraph of $G$ is a graph $G_S$ whose vertex set $V_S$ is a subset of $V$ ($V_S\subseteq V$), and whose edge set $E_S$ is a subset of $E$ ($E_S\subseteq E$).  
A \textbf{query graph} $G_Q = (V_Q, E_Q, L^v_Q, L^e_Q)$ as shown in Figure~\ref{fig:graphs}(e) is usually a smaller graph which represents the pattern of interest. 
A \textbf{root query node} is normally the most selective node in the query graph, which is the starting point of the matching process, that is, $u_0$ in Figure~\ref{fig:graphs}(e). 
A \textbf{query tree} represents a variant of the spanning tree of the query graph, e.g., a Breadth-First Search (BFS) tree shown in Figure~\ref{fig:graphs}(f). 
Here, all the \textbf{tree edges} $(u_p, u)$ connect a parent node $u_p$ to a node $u$. Note that the parent child relationship is not dependent on the direction of edges, e.g., $u_0$ is the parent of $u_2$ even though the edge is directed from $u_2$ to $u_0$. 
The edges not in the query tree are referred to as \textbf{non-tree edges}, e.g, $(u_2, u_5)$. 
A \textbf{matching order} is the  sequence of query nodes and consequently, query tree edges to follow when finding an embedding. 
The non-tree edges are manually verified as we follow the matching order.

\subsection{Problem Definition}
Subgraph matching is the task of listing all the subgraphs of $G$ that match the query graph $G_Q$, i.e., a set $S(G, G_Q)$. \textbf{Real-time subgraph matching} requires continuously computing all the subgraphs that can be found up to the given point in time $t$. 
Specifically, given the data graph $G$ and query graph $G_Q$ at time $t_0$, if a batch of update $\Delta G = \{e_1, e_2, e_3, ..., e_n\}$ is made between $t_0$ and $t_1 (= t_0 + \delta t)$, the system needs to compute a result set $\Delta S$  at $t_1$, such that $S(G\oplus \Delta G, G_Q) = S(G, G_Q) \oplus \Delta S$. 
Here $\Delta S$ represents all the embeddings newly formed (or removed) due to the batch of insertions (or deletions) represented by  $\Delta G$.

Whether or not a subgraph $G_S$ is a match to query $G_Q$ depends on the underlying matching rule. For example, \textbf{isomorphism}~\cite{Han:2013:TIT:2463676.2465300,bi2016efficient,bhattarai2019ceci} requires the embeddings to have a bijective mapping with the query graph. Formally, a subgraph $G_S$ of a data graph $G$ is isomorphic to the query graph $G_Q$ if and only if there exists a bijective function $f: V_Q \rightarrow V_S$ such that $\forall u \in V_Q$, $L^v_Q(u) \subseteq L^v(f(u))$, $\forall(u_i, u_j) \in E_Q$, $(f(u_i), f(u_j)) \in E_S$ and $L^e_Q(u_i, u_j) = L^e(f(u_i), f(u_j))$.
In Figure~\ref{fig:graphs}(b), the graph snapshot $G$ contains two isomorphic embedding of the query graph in Figure~\ref{fig:graphs}(e), where the query edges \{($u_0, u_1$), ($u_2, u_0$), ($u_0, u_5$), ($u_1, u_3$), ($u_1, u_4$), ($u_2, u_6$), ($u_2, u_5$)\} are matched with data edges \{($v_1$, $v_3$),($v_4$, $v_1$),($v_1$, $v_5$),($v_3$, $v_6$),($v_3$, $v_7$),($v_4$, $v_8$),($v_4$, $v_5$)\} respectively on the first embedding while the next embedding matches query edge $(u_2, u_6)$ to data edge $(v_4, v_0)$. Isomorphism is useful in applications that require strict matching such as in chemical compound or protein-network.

Some works utilize looser restrictions on matching such as homomorphism~\cite{kim2018turboflux} and simulation~\cite{ma2011capturing,fan2013incremental}\footnote{Due to lack of space, we refrain from their definition on this paper but we encourage interested readers to refer to the cited works.}. Other variants include the temporal ordering of the edges~\cite{li2019time}, automorphism removal~\cite{bhattarai2019ceci,shao2014parallel}, and approximate matching~\cite{tian2008tale}. All of these solutions first find matches of individual edges(nodes) and combine them while confirming to the desired level of structural similarity. Using the {\matcher} and {\enumerator} functions in API, a user can implement all these variants and many more without having to redesign the whole system stack as we demonstrate in our evaluation.

\vspace{-2mm}
\subsection{Related Works}
\label{sec:related}

The gather-scatter models used by graph computing systems~\cite{malewicz2010pregel} or even the coarse-grained filter-expand models used by graph mining-systems~\cite{teixeira2015arabesque} are not effective for subgraph matching. Therefore, subgraph matching has been subject of active research for the last 2 decades. These works normally only support static graphs and focus on developing number of techniques to optimize the matching order~\cite{bi2016efficient,zhao2010graph},  prune the false candidates~\cite{shang2008taming,zhao2010graph}, optimize index structure~\cite{He:2008:GQL:1376616.1376660,Han:2013:TIT:2463676.2465300},  reduce the data redundancy~\cite{Ren:2015:EVR:2735479.2735493}, search for the multiple queries together~\cite{ren2016multi}, and parallelization~\cite{lai2015scalable,bhattarai2019ceci}.
The way these systems could work for real-time subgraph matching is getting a graph snapshot at a point in time, and running subgraph matching on that snapshot. However, when the interval between two snapshots gets smaller, the redundancy plagues their performance. 

TurboFlux~\cite{kim2018turboflux} has made a breakthrough in incremental subgraph matching using a data-graph centric index structure. While this improves the index update costs, its inability to represent event context, strictly sequential processing, and in-memory index storage makes its utility limited.
A recent work~\cite{li2019time} designs a sliding window, time-constrained subgraph isomorphism, which handles the constant inflow and outflow of edges from the search space in batches. 
However, it stores partially materialized embeddings in the tree, which comes with a very high memory cost and embedding update costs.
{\name} is closest to these works as it provides an incremental approach to filtering-enumeration subgraph matching in streaming graphs. In contrast to the problem-driven approach in these works, it designs a wide-view system that can be customized to different embedding definitions, while providing the system level capabilities to support subgraph matching on large data and query graphs over long run.

\textbf{Join based approaches} treat graph query as relational multi-way join, where the result set is generally expanded one query edge at a time~\cite{choudhury2015selectivity,sun2012efficient} or one query node at a time~\cite{ngoworstcaseoptimal2012,ammar2018distributed} using different join plan~\cite{lai2019distributed}. While they usually are better suited for streaming graphs, provide superior join plan, and work well with parallelization frameworks, they struggle to work for larger query graphs because they perform aggressive expansion of partial matches using node label and/or edge label filters. In contrast, {\name} utilizes the topology of a query graph to perform much better filtering and pruning before expanding the partial embeddings.

\begin{figure}[t]
	\begin{center}
		\includegraphics[scale=0.35]{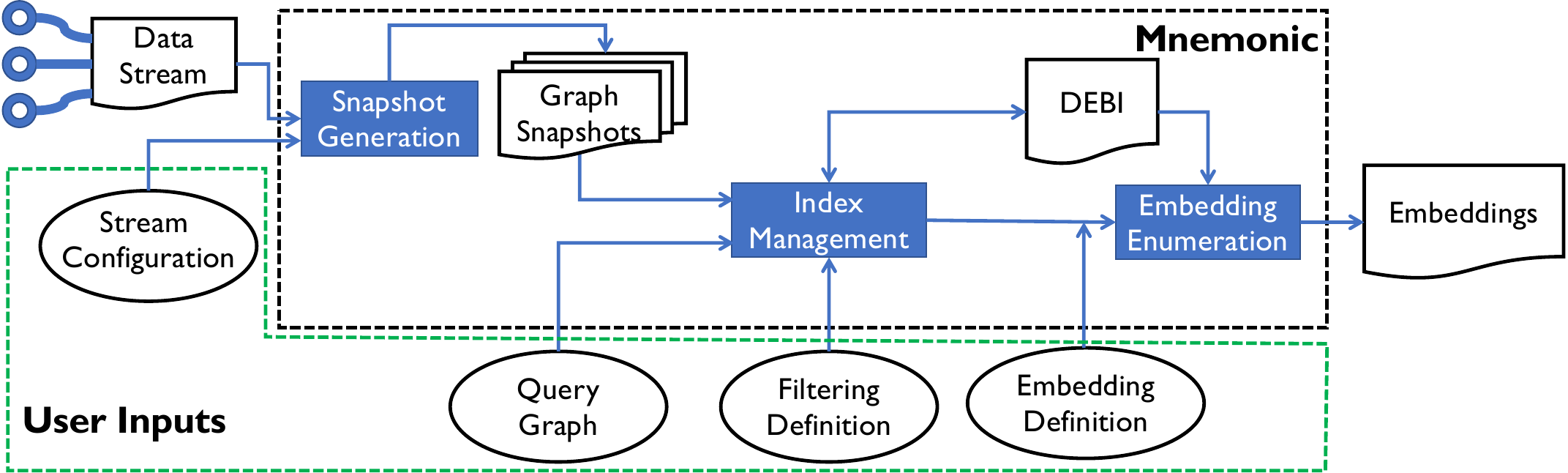}
		\caption{{\name} system overview}
		\vspace{-0.2in}
		\label{fig:overview}
		\vspace{-0.1in}
	\end{center}
\end{figure}
\vspace{-2mm}
\section{{\name}}
\label{sec:overview}

Figure~\ref{fig:overview} illustrates the main components of the {\name} system.
Given a data stream, a user provides four inputs (ovals in Figure~\ref{fig:overview}) to the system - \textit{query graph}, \textit{stream configurations} that are used to customize the snapshot generation, a function {\matcher} that describes the matching conditions between a data and query edge pair, and {\enumerator} that specifies restrictions on the structural properties of an embedding. 
{\name} follows a filtering and enumeration paradigm, where {\matcher} and {\enumerator} helps to customize the two steps respectively.
\begin{algorithm}[h]
	\begin{footnotesize}
		\textbf{input}: queryGraph $G_Q$, dataStream $G$, config\\
		\textbf{output}: set $S(G, G_Q)$ = $\{ \}$ of embedding\\
		
		\tcp{System Initialization}
		\blue{stream\_handle\_t} streamHandle = InitializeStream(dataStream, config);\\
		\blue{index\_handle\_t} indexHandle = InitializeIndex(queryGraph, config);\\
		
		\tcp{Continuous Enumeration of Embeddings}
		\blue{snapshot\_t} s;\\
		\While {(s = getSnapshot())}
		{
			\If{s.insertBatch.size() > 0}
			{
			batchInserts(s.insert\_list, {\matcher}, {\enumerator});\\
			}
			\If{s.deleteBatch.size() > 0}
			{
			batchDeletes(s.delete\_list, {\matcher}, {\enumerator});\\
			}
		}
		\caption{ \footnotesize Subgraph Matching using {\name}}
		\label{algo:mainalg}
	\end{footnotesize}
\end{algorithm}

Under the hood, the system is facilitated by two major data structures and three major components. As specified, the graph is stored as an adjacency list, where each node has its own list of incoming and outgoing edges. The attribute store indexed by node and edge ids store the vertex and edge attributes. The index structure {\idx} is stored alongside edge attributes. 
The snapshot generator generates the new batch of inputs to be processed at a given time.
The index management unit incrementally updates the {\idx} for a given snapshot(Section~\ref{sec:idxmgmt}). 
The embedding enumeration unit prepares the current batch of updates for enumeration, computes matching order and ensures the correctness of result (Section~\ref{sec:enumeration}).

{\name} implements a subgraph matching system as outlined in Algorithm~\ref{algo:mainalg}. First, using data and query based heuristics, it sets up hyper-parameters. This step is normally referred to as preprocessing in existing works~\cite{sun2019inmemory}, which involves tasks such as finding the root query node, generating the query tree, and determining the matching order. The default heuristics for these tasks are chosen in {\name} based on common practices. For example, the most selective node is chosen as the root node, the BFS tree rooted at that node is the spanning tree, and BFS traversal order is the matching order. These tasks can be accomplished  by invoking \emph{InitializeIndex}. \emph{InitializeStream} sets up the snapshot generator, where a few customizing parameters such as \textit{window size}, \emph{stride or batch size} are specified by the user. In each iteration, the last snapshot of the graph and changes made since then are provided with two lists- insertion and deletion list. For example, at time $t2$ in Figure~\ref{fig:graphs}(d), the insertion list \{$(1, 2, 0)$\}, and deletion list \{$(3,7,1), (1,5,0)$\} are generated on top of snapshot $G1$. Once the hyper-parameters are determined, it fetches the new snapshot, makes changes to the {\idx} with the help of {\matcher}, and outputs newly formed (or removed) embeddings from insertions (or deletions) made in the current snapshot by traversing the {\idx}, repeatedly with the help of {\enumerator} as outlined in Algorithm~\ref{alg:batching}.

\begin{figure}[t]
	\begin{center}
		\vspace{-0.1in}
		\includegraphics[scale=0.54]{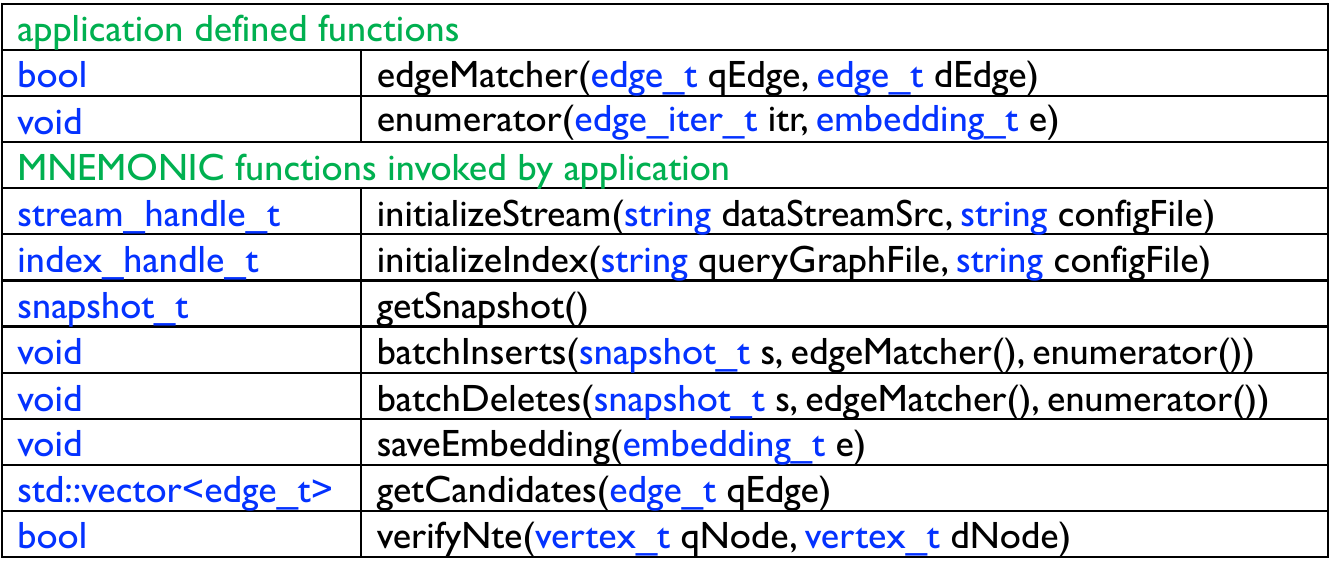}
		\vspace{-0.1in}
		\caption{{\name} API}
		\vspace{-0.2in}
		\label{fig:api}
	\end{center}
\end{figure}
The {\name} API consists of two groups of functions as listed in Figure~\ref{fig:api}. 
First, there are two user-defined functions: {\matcher} and {\enumerator} discussed earlier.
Second, a number of system functions are implemented to perform other essential tasks on behalf of users. 
For example, \emph{initializeStream}() and \emph{initializeIndex}() initialize the graph stream and configure index handle respectively. The function \emph{getSnapshot}() provides access to the latest stable snapshot of the data graph and changes made since then.
\emph{batchInserts}() updates the graph  and {\idx} with insertions and finds newly formed embeddings, while \emph{batchDeletes}() does the same for deletions. Function \emph{getCandidates}() enables the {\enumerator} to fetch the candidates of a given query edge from {\idx}. 
The function \emph{verifyNte}() is used to verify the presence of non-tree connections for the current data-query node pair on partially materialized embedding.



\begin{figure}[t]
	\begin{center}
		\vspace{-0.1in}
		\includegraphics[scale=0.38]{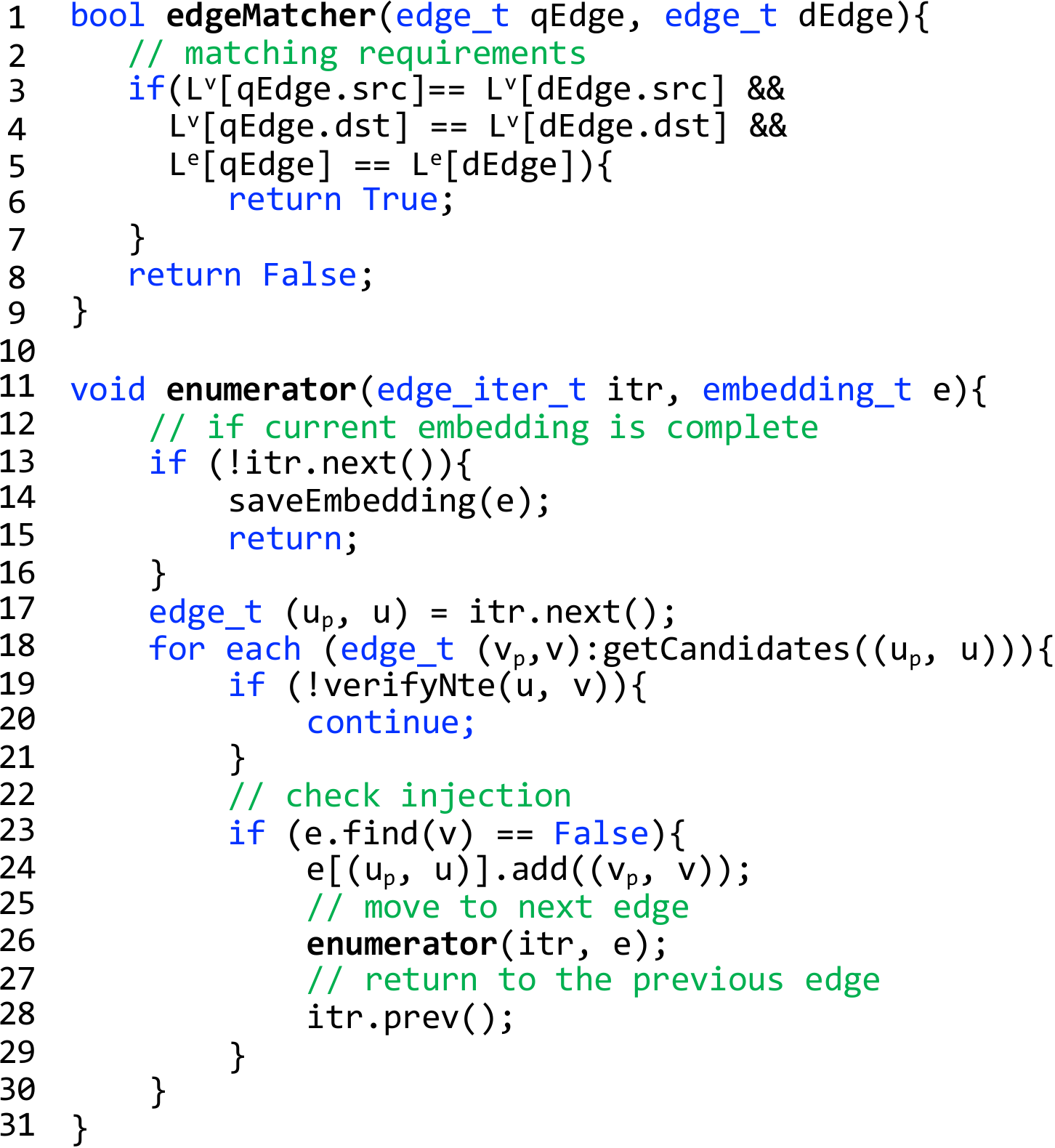}
		\vspace{-0.1in}
		\caption{Implementation of isomorphism on {\name}.}
		\vspace{-0.2in}
		\label{fig:isomorphism}
	\end{center}
\end{figure}

Let us implement {\matcher} and {\enumerator} for \textbf{subgraph isomorphism} using {\name} APIs. Given a data-query edge pair, {\matcher} returns \emph{True} if the edge labels and respective node labels of both endpoints are identical as shown in Figure~\ref{fig:isomorphism} (line 3-5). 
The {\enumerator} for isomorphism (line 11-31) implements a backtracking algorithm that joins matches of individual query edges to build the embeddings. It takes an incomplete embedding \textit{e} and edge iterator \textit{itr} as input. For each edge in the query tree, it retrieves the candidate matches by calling \textit{getCandidates}. The corresponding entry of the embedding e is then set to each of the candidate matches one at a time, while the process is repeated recursively for the next query edge. 
Line 23 makes sure that each node has a distinct match in a given embedding (injective constraint) while line 19 verifies the non-tree edge connections for the pair of $v$ and $u$. When the match for all edges is found, the embedding is written to the result file using function \emph{saveEmbedding} and the backtracking process returns to the previous edge (line 13-16).

By customizing these functions, a user can implement different variants of subgraph matching solutions. For example, the {\enumerator} for \textbf{homomorphism} can be obtained by removing the injective constraint verification from isomorphism, i.e., line 23 from Figure~\ref{fig:isomorphism}. In homomorphism, a single data edge can be used as a match for multiple query edges. \textbf{Dual simulation}~\cite{ma2011capturing} can simply be obtained by joining candidates of each edge and verifying duality for non-tree edges. Similarly, \textbf{strong simulation} can be obtained by adding locality constraint on the {\enumerator} for dual simulation. We skipped the implementation detail of these algorithms due to space constraints, but our evaluation demonstrates {\name}s' ability to program different variants.

\vspace{-2mm}
\section{Index Design}
\label{sec:simu}
\begin{figure}[t]
	\begin{center}
		\vspace{-0.1in}
		\includegraphics[scale=0.3]{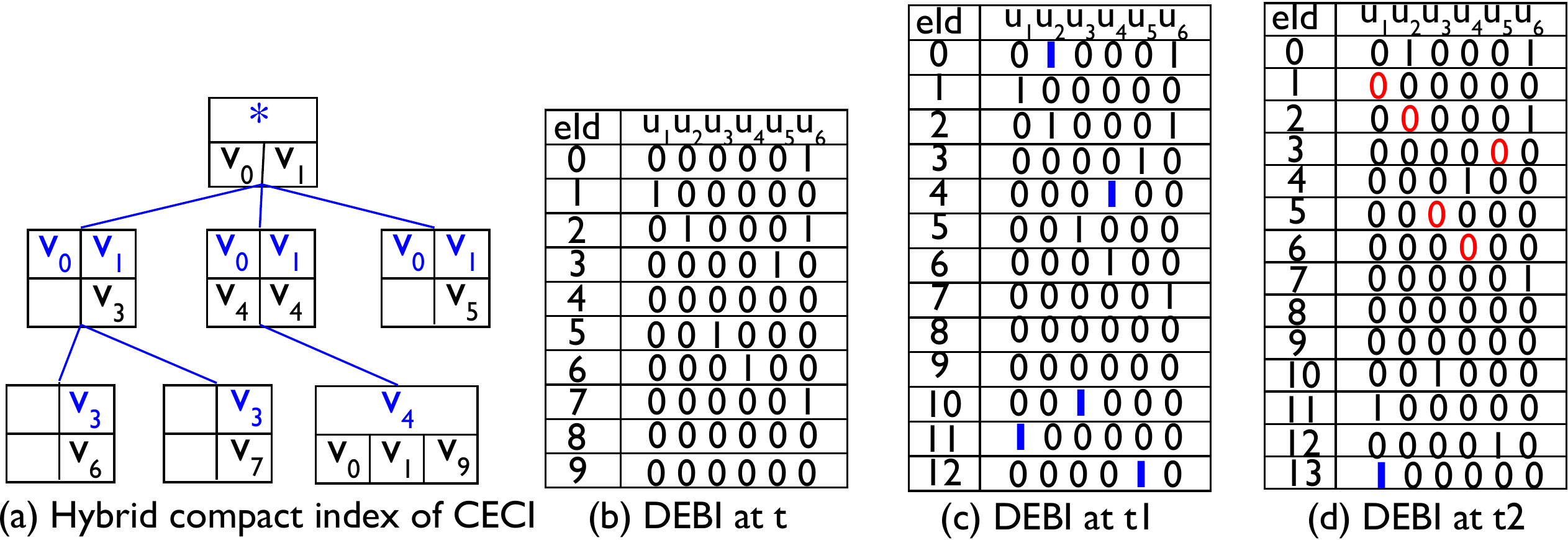}
		\caption{Query-centric candidate store vs data graph centric index of {\name} for the example in Figure~\ref{fig:graphs}. 1 means corresponding data-query edge are match of one-another. }
		\label{fig:dcivsqci}
		\vspace{-0.2in}
	\end{center}
\end{figure}

For the filter and enumerate techniques, naive storage designs such as per-node candidate list or match matrix perform poorly during enumeration as they lose graph topology and have to repeatedly verify edges by searching on the data graph. 
Recent works~\cite{bi2016efficient,bhattarai2019ceci} adopt an approach where explicit candidate lists are collected and stored along with the graph topology. Since it stores the matches of each query node, and their connections to the matches of the neighboring query nodes, search space becomes very compact. Figure~\ref{fig:dcivsqci}(a) shows the index constructed by CECI~\cite{bhattarai2019ceci} for the data graph in Figure~\ref{fig:graphs}(b). Each node in the query tree contains a key-value store, where the keys $v_p$ are the matches of the parent query node $u_p$ and the values are the list of nodes $\{v\}$ that match $u$ and are adjacent to $v_p$.In addition to reducing search space, they obtain better access throughput since the matches of a node are stored consecutively.

\textbf{Observation \#1: Query-centric index structures are not suited for streaming graphs due to high update cost.} In $G2$, when the edge $(v_3, v_{7})$ is deleted, since, it matches the edge $(u_1, u_4)$, one first needs to find the key $v_3$ in the key-value store of $u_4$, and remove $v_{7}$ from entry $ceci[u_4][v_3]$. Locating the key $v_3$ and removing $v_{7}$ both take up to $\mathcal{O}(|V|)$ time.
%
To reduce the update cost, TurboFlux~\cite{kim2018turboflux} proposes a data graph centric index structure known as \emph{dcg}. They design \emph{dcg} as a complete multigraph where every vertex pair $(v_i, v_j)\in E$  has $|V_Q|-1$ edges. Each edge has a query vertex $u$ as the edge label, and its state is one of \emph{Null}, \emph{Implicit}, or \emph{Explicit} based on local and neighborhood attributes. When an edge $(v_i, v_j)$ is inserted to the graph, it simply adds edges to the \emph{dcg}. 

\textbf{Observation \#2: While Turboflux is easier to update and enables the incremental subgraph matching, it fails to preserve event context and even produces false positives.} 
In \emph{dcg}, all instances of an edge between two endpoints are considered the same. While they have the option to re-enumerate all embedding involving an edge when it is inserted again, it fails to distinguish between 2 instances of the edges. However, such distinction is very important in many applications.  For example, in cyber forensic, a login to the server by a user after compromise is very significant to attack impact analysis compared to one before compromise. Including the benign login in the embedding would produce a false positive. Uniquely identifying such events is necessary in order to generate useful embeddings.



\vspace{-2mm}
\subsection{{\idx}}
\label{sec:computemodel}
Each edge in the data graph has a unique identifier associated with it henceforth referred to as \emph{edgeId}. 
Figure~\ref{fig:dcivsqci}(b) shows the {\idx} for graph snapshot $G$, while Figure~\ref{fig:dcivsqci}(c)-(d) shows the updated {\idx} at time $t1$ and $t2$ respectively after processing two snapshots.
{\idx} maintains a bitmap of size $|V_Q|-1$ bit for each data graph edge, where each bit represents whether the data edge matches the corresponding query tree edge or not. The information about whether an edge $(v^{\prime}, v)$ with the id of \emph{edgeId} matches $(u_p, u)$ is stored at $\idx[\emph{edgeId}][u]$. 
In addition, the matching nodes of the root query node ($u_0$) are stored in a bit vector of size $|V|$, known as {\idxroots}. The size of {\idx} is equal to $(|E| \times (|V_Q|-1) + |V|)$ bits, which is same as that of TurboFlux~\cite{kim2018turboflux} and smaller than $\mathcal{O}(|E_Q|\times|E|)$ of CECI~\cite{bhattarai2019ceci}, $\mathcal{O}(|E|\times |V_Q|)$ of CFLMatch~\cite{bi2016efficient}, and $\mathcal{O}(|E|^{|V_Q|})$ of TurboIso~\cite{Han:2013:TIT:2463676.2465300}.

\noindent\textbf{Updating:} 
The time for inserting, updating, or reading the entry for any data and query edge pair in {\idx} is constant, i.e., $\mathcal{O}(1)$.
When an edge $(v^{\prime}, v)$ with the id of \emph{edgeId} is inserted, the entry $\idx[\emph{edgeId}]$ is initialized. For all $(u_p, u)$ in the query tree that matches $(v^{\prime}, v)$, $\idx[\emph{edgeId}][u]$ is set to \emph{True}. If the inserted edge matches multiple query edges, all the entries that need to be changed are stored together in {\idx}. Once the initial edges to change for are identified, the graph is traversed downward and upward to reflect the changes of update in the {\idx} as explained in Section~\ref{sec:idxmgmt}.

{For the graph in Figure~\ref{fig:graphs}(b) and index in Figure~\ref{fig:dcivsqci}(c), when $(v_0, v_2)$ is inserted, $\idx[11][u_1]$ is updated to \emph{True}. In addition, $\idx[10][u_3]$ and $\idx[4][u_4]$ are now changed to \emph{True}. 
 Since $(v_4, v_1)$ matches both $(u_2, u_0)$ and $(u_2, u_6)$, we need to set $\idx[2][u_2]$ and $\idx[2][u_6]$ to $True$.
In contrast, CECI in Figure~\ref{fig:dcivsqci}(a) needs to first find $v_0$ in $CECI[u_1]$ and add $v_2$ to it. Then, one needs to repeat the process by searching $v_2$ at $u_3$ and $u_4$, and adding $v_6$ and $v_7$ to the respective lists, which costs $\mathcal{O}(|V|)$ each. }

\noindent \textbf{Accessing:} 
{\idx} achieves an equivalency to CECI~\cite{bhattarai2019ceci} in terms of accessing candidates. 
Given a query edge $(u_p, u)$, \emph{getCandidates} function returns all edges $(v_p, v)$ with the id of \emph{edgeId} such that $\idx[\emph{edgeId}][u]$ is \emph{True}, where $(*, v_p)$ is already matched with $(*, u_p)$. Since all outgoing/incoming edges of $v_p$ are stored together in the adjacency list, all the candidates can be fetched in one read. A small penalty is paid to filter out the edges that do not match $(u_p, u)$. 

\noindent\textbf{Memory recycling:}
In {\name}, the memory space used by deleted edges is recycled for upcoming edges.  When an edge is deleted, 
{\name} first locates this edge in the adjacency list of the corresponding vertex and swaps it with the last entry of the adjacency list, and reduces the degree of the vertex. 
When a new edge for the same vertex is inserted later, the degree is increased and the \emph{edgeId} of the previously deleted edge is reused. When no \emph{edgeId} is available, a new entry is inserted. In Figure~\ref{fig:graphs}(d), when the edge $(v_1, v_{9})$ is inserted after $t2$, it uses the id of the last deleted edge of $v_1$, i.e. $edgeId=3$ of $(v_1, v_5)$. By reclaiming the memory from deleted edges and their corresponding {\idx} entries, {\name} is able to keep index size in check.


\noindent \textbf{External memory support:}
{\name} provides an option to backup edges and their attributes to the external-memory storage when search space is large. It adopts a simple FIFO retention policy, where the newly generated events are kept in the memory. The amount of edges to be kept in memory is specified by a user-controlled variable called \textit{in-memory window}. All the edges $(v_p, v)$ with an id of \emph{edgeId} outside of this window are added to a buffer along with $\idx[edgeId]$.  When the buffer is full, it is dumped into the disks. The vertex information is maintained entirely in memory, while the edges on the disk are stored using transactional edge logs~\cite{zhu2019livegraph}, so that the adjacency list of a given node can be fetched in a single transaction.
\vspace{-2mm}
\section{Index Management and Update}
\label{sec:idxmgmt}
\vspace{-2mm}



\begin{algorithm}[t]
	\begin{scriptsize}
		\DontPrintSemicolon
		\SetKwBlock{Begin}{function}{end function}
		\tcp{Processing the batch of insertions}
		\Begin(batchInserts {($batch,{\matcher},{\enumerator}$)}) 
		{
			updateGraph($batch$);\\
			$frontier$ = getFrontier($batch$, {\matcher});\\
			topDownFiltering($frontier$);\\
			bottomUpFiltering($frontier$);\\
			enumeration({\enumerator} );\\
		}
		\tcp{Processing the batch of deletions}
		\Begin(batchDeletes {($batch, {\matcher}, {\enumerator}$)}) 
		{
			$frontier$ = getFrontier($batch$, {\matcher});\\
			enumeration({\enumerator} );\\
			updateGraph($batch$);\\
			bottomUpFiltering($frontier$);\\
			topDownFiltering($frontier$);\\
		}
		\caption{Batch Processing }
		\label{alg:batching}
	\end{scriptsize}
\end{algorithm}

When an edge is inserted to or deleted from the data graph, {\idx} needs to be updated. 
Given a data edge $(v_p, v)$ with an id of \emph{edgeId}, and a query tree edge $(u_p, u)$, the entry $\idx[edgeId][u]$ is set \emph{True} when the following two conditions are met:
(1) (\textbf{edge match}) The data edge $(v_p, v)$ matches $(u_p, u)$ as defined by the {\matcher}, and  
(2) (\textbf{neighborhood match}) For each neighbor $u^{\prime}$ of $u$, there is at least one edge $(v, v^{\prime})$ that matches the query edge $(u, u^{\prime})$. Similarly, for each neighbor $u^{\prime}_p$ of $u_p$, the edge $(u_p, u^{\prime}_p)$ has at least one edge $(v_p, v^{\prime}_p)$ that is a match.

For each edge insertion or deletion, a {\idx} entry needs to be initialized or cleared. In addition, each update in the data graph may change the {\idx} for many neighboring edges.
The effect propagates up to $d$-hop away from the initial edge, where $d$ is the diameter of the query graph.  
Multiple iterations through this neighborhood are required before an equilibrium is reached and the minimal candidate set is obtained, which is shown to be NP-Hard~\cite{bi2016efficient}. {\name}, similar to existing works, relies on single pass top-down and bottom-up filtering following the matching order. After the completion, this process verifies the duality of every candidate of each edge in the query tree. The matches for edges not in the query tree, that is, \textit{non-tree edges}, are verified during enumeration. 

 TurboFlux~\cite{kim2018turboflux} also performs a downward exploration and upward traversal from the new edge. Once it completely updates the \emph{dcg} for an edge and enumerates the new embeddings involving that edge, it moves to another edge. The traversal cost for each edge insertion or deletion can go as high as $\mathcal{O}(|E|*|V_Q|)$. While they can parallelize the traversal for each edge update, they need to synchronize after each step downward or upward. Since each step has only a limited amount of  work, they fail to take advantage of multi-core machines. 

\vspace{-2mm}
\subsection{Batch Processing}
\label{sec:batching}
\vspace{-2mm}
Batching the edges into subgrapgh matching system needs to solve two problems. First, independently parallelizing the top-down and bottom-up traversal used by Turboflux yields wrong result due to out of order processing of edges. Second, huge amount of redundancy may exist between the traversal space of different edges, and traversing them for each edge update is not efficient.
For example, in snapshot $G1$ from Figure~\ref{fig:graphs}(c), the traversal space for $(v_2, v_6)$ and $(v_0, v_2)$ overlap significantly. Specifically, the sub-tree rooted at $v_2$ needs to be traversed for both edge insertions.


To solve these problems, we use a data structure called \textbf{unified traversal frontier}. Initially, this frontier for each edge $(u_p, u)$ contains all the edges $(v_p, v)$ with id \emph{edgeId} from the inserted edges that match $(u_p, u)$ as defined by the {\matcher}. 
When the effect of edge insertion/deletion propagates during exploration, the frontier entries are updated with the affected edges which matches the respective query edge. In essence, the unified traversal frontier represents the area in the graph affected by the current batch of updates.
In the snapshot $G1$ from Figure~\ref{fig:graphs}(c), the initial traversal frontiers for edges $(u_0, u_1)$, $(u_1, u_3)$, and  $(u_0, u_5)$ are $\{(v_0, v_2)\}$, $\{(v_2, v_6)\}$, and $\{(v_0, v_5)\}$ respectively.

\textbf{Batching Insertions:} Algorithm~\ref{alg:batching} shows the order of steps in ingesting a batch of insertion. The \textbf{top-down filtering} traverses the data graph following the BFS traversal order of the query tree and filters out the candidates that do not match the current edge.
Given a data graph edge $(v_p, v)$ with an id \emph{edgeId} and a query tree edge $(u_p, u)$, $\idx[edgeId][u]$ is set \emph{True} if  
(1) \textbf{f1}: there is a node $v_s$ in the data graph such that the path from $v_s$ to $v_p$ matches the path from $u_0$ to $u_p$, 
(2) \textbf{f2}: if the query node $u$ has $n_l$ incoming edges with label $l$, the data node $v$ must have at least $n_l$ incoming edges of label $l$, and 
(3) \textbf{f3}: if the query node $u$ has $n_l$ in-neighbors with label $l$, the data node $v$ must have at least $n_l$ in-neighbors of label $l$. Both \textbf{f2} and \textbf{f3} need to be true for outgoing edges and neighbors as well. 

For a query edge $(u_p, u)$, the frontier consists of all the edges from current batch that matches it, and all the edges adjacent to $v_p$, where $(*, v_p)$ matches $(*, u_p)$(\textbf{f1}). For each edge $(v_p, v)$ in the frontier with id \emph{edgeId}, if $\idx[edgeId][u]$ is \emph{True} or if there is another edge $(v_p^{\prime}, v)$ (id of $edgeId^{\prime}$) such that $\idx[edgeId^{\prime}][u]$ is \emph{ True}, the $\idx[edgeId][u]$ is set \texttt{True} and the edge is removed from frontier. If edge $(v_p, v)$ and $(u_p, u)$ do not meet conditions specified in \textbf{f2} or \textbf{f3} it is removed from frontier. Otherwise, $\idx[edgeId][u]$ is set \texttt{True}.
In snapshot $G1$, since $(v_0, v_2)$ satisfies \textbf{f2} and \textbf{f3} for $(u_0, u_1)$, $\idx[11][u_1]$ is set \emph{True}. Later, since edges $(v_3, v_6)$ and $(v_3, v_7)$ are already matches of $(u_1, u_3)$ and $(u_1, u_4)$, $\idx[10][u_3]$ and $\idx[4][u_4]$ are readily set \emph{True} and are removed from the corresponding frontiers. Here, 10 and 4 are id of $(v_2, v_6)$ and $(v_2, v_7)$ respectively.

The \textbf{bottom-up filtering} (line 5) routine processes one query edge at a time in reverse BFS traversal order. 
Given an edge $(v_p, v)$ with the id of \emph{edgeId} such that $\idx[edgeId][u]$ is \emph{True}, the bottom-up filtering rule \textbf{f4} makes sure that query sub-tree rooted at $u$ matches a tree in data graph rooted at $v$.  In other word, for each child $u_c$ of $u$ in the query tree, there must be an edge $(v, v_c)$ with an id \emph{$edgeId_c$} in data graph such that $\idx[edgeId_c][u_c]$ is \emph{True}. This process is not required for edges with a leaf node as one endpoint.
 For an edge $(v_p, v)$ with id $edgeId$  in the frontier of an edge $(u_p, u)$, if $\idx[edgeId][u]$ is \emph{True} and \textbf{f4} is not satisfied, $\idx[edgeId][u]$ is changed to \emph{False}. Now,  all the edges $(*, v_p)$ with an id $edgeId_p$ such that $\idx[edgeId_p][u_p]$ is \emph{True} are added to the frontier of parent query edge $(*, u_p)$.
In snapshot $G1$, since $(v_2, v_6)$ and $(v_2, v_7)$ are matches of $(u_1, u_3)$ and $(u_1, u_4)$ respectively, $\idx[11][u_1]$ remains \emph{True}. Similarly, for a node $v$ in the frontier of $u_0$, if \textbf{f4} is not satisfied, {\idxroots[v]} is changed to \emph{False}.  Since $v_0$ now has matching neighbors for each of $u_1$, $u_2$, and $u_5$, {\idxroots[$v_0$]} remains \emph{True}.

\textbf{Batching Deletions:} 
The process for the batch of deletion is somewhat reverse to the insertion. Starting from the deleted edges, the enumeration is performed on {\idx} before updating. Later the batch of deleted edges are removed from the graph and the top-down filtering follows the bottom-up in order to update the {\idx}. The frontier for each query node $(u_p, u)$ initially contains the edges (id = \emph{edgeId}) from the deletion batch such that $\idx[edgeId][u]$ is \emph{True}.  
During \textbf{bottom-up exploration}, for each edge $(v_p, v)$ with id \emph{edgeId} in the frontier of the query edge $(u_p, u)$, $\idx[edgeId][u]$ is set to \emph{False}. If the pair $v_p$ and $u_p$ does not satisfy \textbf{f4}, all edges $(*, v_p)$ with an id $edgeId_p$ such that $\idx[edgeId_p][u_p]$ is \emph{True} are added to the frontier of $(*, u_p)$. For the root query node, the {\idxroots} is updated using the same rule. 
In the snapshot $G2$, when an edge $(v_3, v_7)$ (id = 6) is deleted, $\idx[1][u_1]$ is changed to \emph{False}, i.e., $(v_1, v_3)$ no longer matches $(u_0, u_1)$. Later, when $(v_1, v_5)$(id = 3) is deleted, the pair $v_1$ and $u_0$ does not satisfy \textbf{f4} as $v_0$ does not have a neighbor matching $u_5$. Thus, $\idxroots[v_1]$ is set to \emph{False}.

\begin{figure*}[t]
	\begin{center}
		\vspace{-0.1in}
		\includegraphics[scale=0.52]{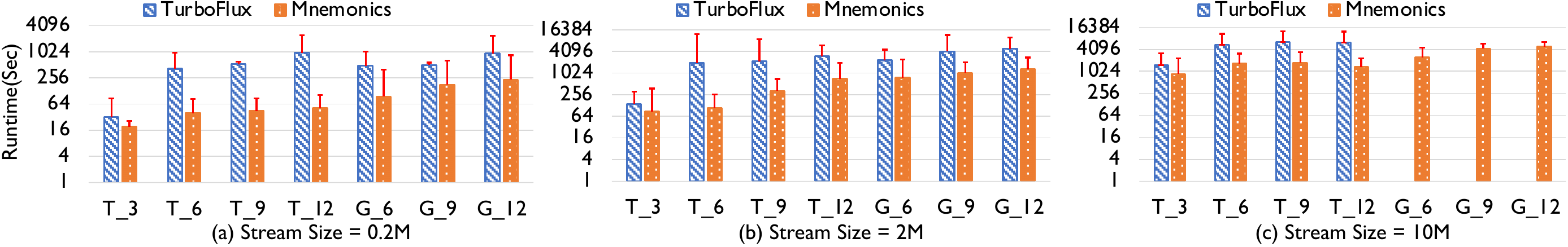}
		\vspace{-0.1in}
		\caption{Runtime of TurboFlux and {\name} on {\netflow}}
		\label{fig:vstf_netflow}
		\vspace{-0.25in}
	\end{center}
\end{figure*}

During {top-down filtering}, for all $(*, v_p)$ in the frontier of $(*, u_p)$, all edges $(v_p, v)$ with id \emph{edgeId} are obtained, and $\idx[edgeId][u]$ is set \emph{False}. If there is no other edge $(v^{\prime}_p, v)$ with id $edgeId^{\prime}$ such that $\idx[edgeId^{\prime}][u]$ is \emph{True}, $(v_p, v)$ is added to the frontier of $(u_p, u)$. 
In snapshot $G2$, since $\idxroots[v_1]$ is $False$, $\{(v_1, v_2), (v_1, v_3)\}$ are initially added to the frontier of $(u_0, u_1)$. But since $(v_0, v_2)$ still matches $(u_0, u_1)$, $(v_1, v_2)$ is removed from the frontier. However, $(v_1, v_3)$ is kept in the frontier as there is no other edge incident on $v_3$ that matches $(u_0, u_1)$. The effect propagates all the way down and the whole sub-tree rooted at $v_1$ is no longer a match to their counterpart in the query tree.

\vspace{-2mm}
\section{Embedding Enumeration}
\vspace{-2mm}
\label{sec:enumeration}
Embedding enumeration is carried out with the help of a user-defined function {\enumerator}. Under the hood, {\name} needs to do a few tasks to improve the performance and ensure correctness of results.

\textbf{Work decomposition:} For a batch of updates,
{\name} decomposes the incoming batch into fine-grained work units, i.e., data-query edge pair.  Since the real-world graphs follow power-law nature, this finer-grained decomposition can obtain better load balance. For each data edge $(v_p, v)$ with an id \emph{edgeId} inserted to or deleted from the graph, an initial embedding $\{(v_p, v)\}$ is formed if $\idx[edgeId][u]$ is \emph{True}. If the inserted edge $(v_x, v_y)$ matches a non-tree edge $(u_x, u_y)$, then potentially multiple initial embeddings $\{(v^{\prime}, v_y), (v^{\prime\prime}, v_x))\}$ are generated such that these edges match the tree edges $(u^{\prime}_x, u_x)$ and $(u^{\prime}_y, u_y)$ respectively. Each of these initial embeddings represent one work unit and are dynamically distributed among workers using a pull based technique. In the snapshot $G1$, initial embeddings for enumeration are \{$(v_2, v_6)$$\rightarrow$$(u_1, u_3)$\}, \{$(v_0, v_2)$$\rightarrow$$(u_0, u_1)$\}, and \{$(v_0, v_5)$$\rightarrow$$(u_0, u_5)$\}.

\textbf{Matching order computation:} For static graphs, once the matching order is computed, it is used throughout the process. {\name} strictly follows the BFS order of the query tree during filtering. However, since initial embedding $(v_p, v)$ can match any of the query edges $(u_p, u)$, matching order may start from any of the query graph edges. Thus, a different matching order needs to be computed for enumeration starting at every different edge in the query graph. For a tree edge $(u_p, u)$, the path from the node $u$ to the root query node is placed first in the matching order and the BFS traversal order is used for the rest of the query tree. For a non-tree edge $(u_x, u_y)$, the matching order is $\{(u^{\prime}_y, u_y), (u^{\prime}_x, u_x)\}$ followed by the path from $u_x$ to the root and the remaining edges in the BFS order. Here, $u_x^{\prime}$ and $u_y^{\prime}$ are parent nodes of $u_x$ and $u_y$ respectively. Once the initial embeddings $e$ and their corresponding matching order is obtained, multiple instances of {\enumerator} are spawned. 
In snapshot $G1$, the inserted edge $(v_2, v_6)$ matches $(u_1, u_3)$, thus the matching order is \{($u_1$, $u_3$), ($u_0$, $u_1$), ($u_2$, $u_0$), ($u_0$, $u_5$), ($u_1$, $u_4$), ($u_2$, $u_6$)\}. Similarly, for edge $(v_0, v_2)$, the matching order is \{($u_0$, $u_1$),($u_2$, $u_0$), ($u_0$, $u_5$),($u_1$, $u_3$),($u_1$, $u_4$), ($u_2$, $u_6$)\}.

\textbf{Duplicates Removal:} When a batch of insertion is processed, since {\idx} is updated for the whole batch before the enumeration, many duplicates can be generated. Embeddings that use 2 or more edges inserted in the current batch are listed for each new edge. For example, after updating {\idx} to Figure~\ref{fig:dcivsqci}(c) at time $t1$, two new embeddings are obtained, i.e., the query edges \{($u_1$, $u_3$), ($u_0$, $u_1$), ($u_2$, $u_0$), ($u_0$, $u_5$), ($u_1$, $u_4$), ($u_2$, $u_6$), ($u_2$, $u_5$)\} are matched to edges with ids \{ \textbf{10}, \textbf{11}, 0, \textbf{12}, 4, 7, 9\} and \{ \textbf{10}, \textbf{11}, 0, \textbf{12}, 4, 2, 9\} respectively. Same two embeddings are obtained for each of three edge insertions, i.e., $(v_2, v_6)$, $(v_0, v_2)$, and $(v_0, v_5)$ even though they are formed only after inserting the last edge. Similar problem also arises for a deletion batch when removed embedding contains two or more edges on the current batch. 

\begin{table}[t]
	\begin{center}
		\caption{Sample masking table for a query graph.}
		\vspace{-.04in}
		\begin{scriptsize}
			\resizebox{.48\textwidth}{!}{
				\begin{tabular}{|l|l|l|l|l|l|l|}
					\hline
					$(u_0, u_1)$ & $(u_2, u_0)$ & $(u_0, u_5)$ & $(u_1, u_3)$ & $(u_1, u_4)$ & $(u_2, u_6)$ & $(u_2, u_5)$ \\ \hline
					*111111      & 0*11111      & 00*1111      & 000*111      & 0000*11      & 00000*1      & 000000*      \\ \hline
				\end{tabular}
			}
		\end{scriptsize}
		\vspace{-.2in}
		\label{tbl:masking}
	\end{center}
\end{table}


{\name} uses a masking technique to eliminate the duplicates. For each edge in query graph, a different masking array of size $|E_Q|$ bits is used, where the edges with mask bits set cannot use edges in the current batch as their match in embedding. Table~\ref{tbl:masking} shows a sample mask array for edges in our query graph. In snapshot $G1$, the edge $(v_2, v_3)$ when starting at $(u_1, u_3)$ cannot use $(v_0, v_2)$ as match for $(u_0, u_1)$ or $(v_0, v_5)$ as match for $u_0, u_5$.  However, when starting enumeration from $(u_0, u_1)$ for new edge $(v_0, v_2)$, the masks for neither of $(u_0, u_5)$ and $(u_1, u_3)$ are set. Thus we can use $(v_0, v_5)$ and $(v_2, v_3)$ as matches for $(u_0, u_5)$ and $(u_1, u_3)$ respectively.

\vspace{-2mm}
\section{Evaluation}
\label{sec:experiments}
\vspace{-1mm}
{\name} is implemented using \texttt{C++} and \texttt{OpenMp} on Linux, and evaluated on a server with \texttt{Intel Xeon Gold 6126 24-core 2.60GHz CPU} with 512 GB of memory. We demonstrate its ability to support different stream types(Section~\ref{sec:streamtypes}), search granularity(Section~\ref{sec:granularity}), and embedding types(Section~\ref{sec:embeddingtype}). In addition, we evaluate its memory consumption(Section~\ref{sec:memory}), and parallelization(Section~\ref{sec:granularity}). 
We compare against \textbf{four state-of-the-art solutions}: (1) TurboFlux~\cite{kim2018turboflux} and (2) BigJoin~\cite{ammar2018distributed}, two systems for subgraph matching on a streaming graph; (3) CECI~\cite{bhattarai2019ceci}, a system for subgraph isomorphism on the static graph; and (4) Li et al.~\cite{li2019time} for time-constrained isomorphism on the streaming graph \footnote{The binary for TurboFlux was obtained from the author maintained website~\cite{bin_turboflux}, while the source code for BigJoin,  CECI, and Li et al. the corresponding GitHub repositories.}. 

We use \textbf{3 datasets} for experiments.
(1) {\netflow}  contains anonymized passive traffic traces collected from high-speed internet backbone links, with 18,520,759 triplets in total. Each triplet represents the source, destination, and transport layer protocol of a flow event. There is no deletion of edges or vertices on this graph, i.e., an insert-only stream. 
(2) {\lsbench} consists of simulated RDF social network activities for 100K users generated using the Linked Stream Benchmark data generator~\cite{lsbench}.  Each event represents one of many activities such as posts, comments, and locations. 
The stream contains 23,320,426 triplets, out of which the first 20.9M are insertions and 10\% of the remaining 2.3M updates are deletions. The deletions are generated by randomly picking edges from the first 20.9M edges. The deleted edges are indicated on stream by negating the both endpoints, i.e., $(-1, -3, l)$ if edge $(1,3, l)$ is deleted.
(3) {\lanl} contains the stream of NetFlow records collected from network devices inside Los Alamos National Lab~\cite{turcotte2017unified}. The events recorded during the first 3 days are used, 540M  events in total.
These graphs have respectively 1, 1,  and 6 node types and 8, 45, and 3 edge types.
\textbf{Two types of queries} of different sizes are used, i.e., acyclic (tree) and cyclic (graph) queries. We followed the technique adopted by TurboFlux~\cite{kim2018turboflux} for query generation in {\netflow} and {\lsbench} datasets. 
In total, 100 tree queries each of size 3, 6, 9, and 12 (denoted as $T\_3$, $T\_6$, $T\_9$, and $T\_{12}$), and 100 graph queries each of size 6, 9, and 12 (denoted as $G\_6$, $G\_9$, and $G\_12$) are generated. 
The average runtime for 100 queries is reported unless specified otherwise. For {\lanl}, 100 tree and graph queries are generated where the query edges have timestamps from the corresponding edge in the data graph. These queries are used to evaluate the performance of temporal subgraph matching in Section~\ref{sec:embeddingtype}.


\vspace{-2mm}
\subsection{Varying Stream Types}
\label{sec:streamtypes}
\vspace{-1mm}
 


\noindent \textbf{Insertion only stream:} The comparison of runtime between TurboFlux and {\name} for isomorphism search on {\netflow} is presented in Figure~\ref{fig:vstf_netflow}. In order to assess different amounts of updates, three stream sizes of 0.2M, 2M, and 10M are chosen. For each stream size, the remaining edges from the total of 18.5M edges are loaded in the initial graph (or the first snapshot for {\name}). Both systems are set to  timeout after two hours.  {\name} uses a batch size of 16K.  

Figure~\ref{fig:vstf_netflow}(a) shows the average runtime for processing 0.2M edge insertions. {\name} comes on top with an overall average speedup of $7.8\times$ for all cases(maximum $18.9\times$ on $T\_12$, minimum $1.6\times$ on $T\_3$). The main advantage comes from batching of edges (Section~\ref{sec:granularity}), and finer grained enumeration parallelization. 
In 2M and 10M stream sizes, the average speedup  of $5.9\times$ and $3.2\times$ are obtained. At first glance it seems like speedup decreases with the increase in stream size. 
This is due to the skewed statistics from excluding the timed-out queries. For 2M edge updates, eight queries timed out in {\name}, whereas in addition to these eight, another 29 queries timed out for TurboFlux. 
The timed out queries in {\name} are corner cases with a very high number of matches. For 10M edge insertions, none of the graph queries were finished in TurboFlux within 2 hours while 11, 11, 15, and 23 of $T\_3$, $T\_6$, $T\_9$, and $T\_12$ respectively timed out.  In contrast, all tree queries finished in {\name}, while 7, 13, and 16 queries from $G\_6$, $G\_9$, and $G\_12$  timed out respectively.  If only the runtime of queries finished by TurboFlux were included, the average speedup would be $9.5\times$ and $10.2\times$ for 2M and 10M insertions respectively.  The runtime is fairly consistent for most queries, and standard deviation error is mostly due to a few long-running edge cases. 

\begin{figure}[h]
	\begin{center}
		\vspace{-0.1in}
		\includegraphics[scale=0.420]{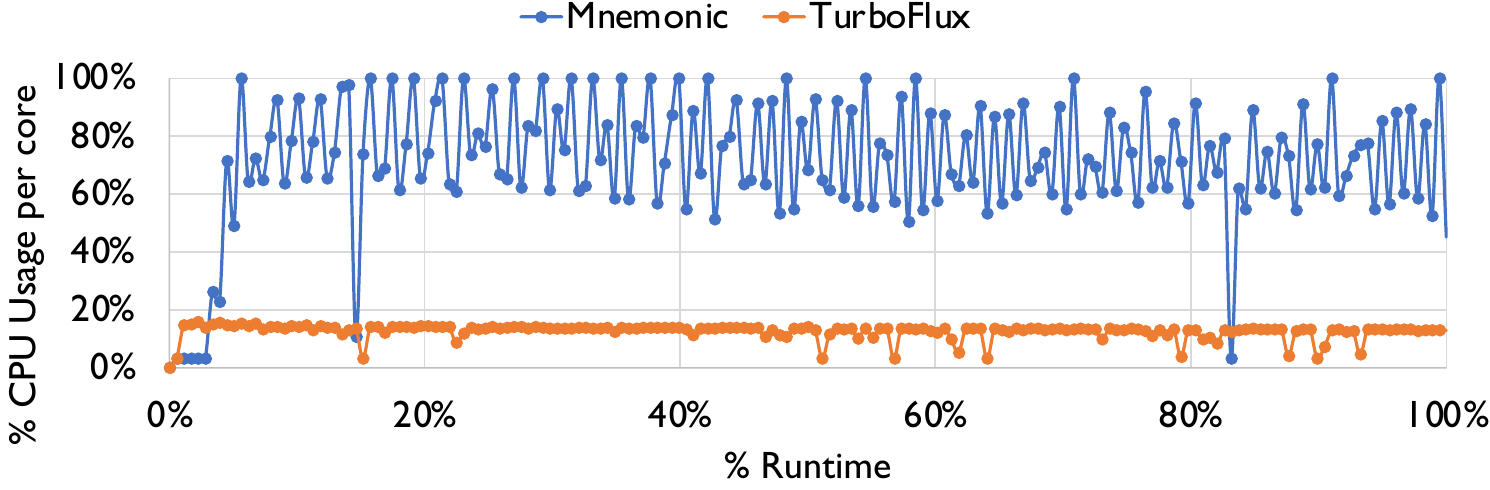}
		\vspace{-0.1in}
		\caption{CPU usage per core over the computation lifetime for query $T\_9 - Q_1$ for {\name} and TurboFlux.}
		\label{fig:cputime}
	\end{center}
	\vspace{-0.2in}
\end{figure}

\begin{figure}[h]
	\begin{center}
		\vspace{-0.1in}
		\includegraphics[scale=0.53]{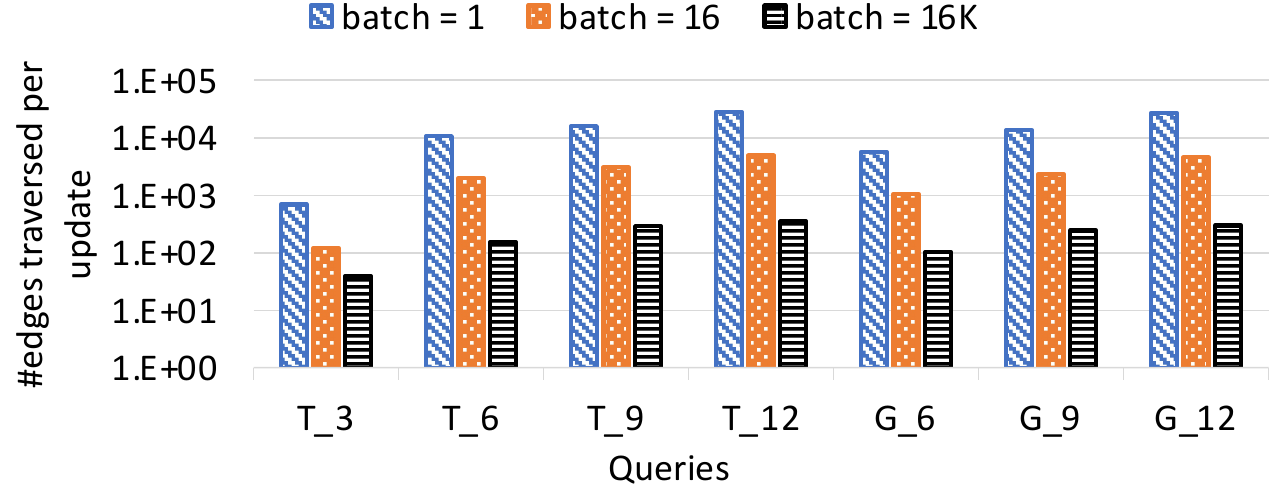}
		\vspace{-0.1in}
		\caption{Traversals per edge update (insertion/deletion) for different batch size. It is the sum of number of edges traversed during top-down and bottom-up filtering.}
		\label{fig:traverse}
	\end{center}
	\vspace{-0.2in}
\end{figure}

Figure~\ref{fig:traverse} compares the average number of edges traversed for reflecting the effect of a single edge update on {\idx}. Since the edges share the traversal within the insertion/deletion batch, the average number of traversals per insert/delete decreases as we increase the batch size. With the increase in the query size number of edges traversed tends to increase as traversal diameter increases. But, this increment follows query size sub-linearly because larger query graphs have more selective nodes and that can eliminate more edges from traversal path.

While the number of traversals directly translates to the time for {\idx} update time, the runtime for embedding enumeration is dependent on the number of embeddings. Since enumerating each embedding is an embarrassingly parallel problem, the batching combined with fine grained work decomposition in {\name} (Section~\ref{sec:enumeration})  keeps all the workers busy during the enumeration phase. Figure~\ref{fig:cputime} shows the cpu usage per core over the runtime for searching a query from $T\_9$ on {\netflow} graph for {\name} and TurboFlux. As  expected, {\name} has consistently better cpu usage over the program's life compared to Turboflux. The two extreme drops are because of the long tails formed where single edge update produces many matches.

BigJoin~\cite{ammar2018distributed} is faster than {\name} and Turboflux for smaller queries. We computed its runtime for homomorphism of some common queries as listed in Table~\ref{tbl:bigjoin}. BigJoin performed worse on rectangle and dual-triangle than 5-cliques because it cannot take as much advantage of intersection to minimize match set in sparse queries. As the query gets bigger, the performance worsens for BigJoin as predicted by~\cite{sun2020rapidmatch}. It takes an average of 6.75 hours to finish $T\_9$ on {\netflow} (stream size = 2M) and longer than 12 hours for $T\_12$, $G\_9$, and $G\_12$, i.e., an order magnitude longer than {\name}.

\begin{table}[h]
	\caption{Query Runtime on {\netflow} (Seconds)}
	\vspace{-0.1in}
	\label{tbl:bigjoin}
	\resizebox{\columnwidth}{!}{
		\begin{tabular}{|l|l|l|l|l|l|}
			\hline
			\multicolumn{1}{|r|}{Queries} & triangle & 4-clique & 5-clique & rectangle & dual-triangle \\ \hline
			BigJoin                       & 43.83   & 167.34 & 293.32 & 356.32  & 325.58      \\ \hline
			TurboFlux                     & 372.38   & 874.67   & 1423.54  & 1657.32   & 1,588.56      \\ \hline
			MNEMONIC                      & 50.46    & 174.48   & 282.83   & 250.09    & 237.26        \\ \hline
		\end{tabular}
	}
\vspace{-0.1in}
\end{table}

\begin{figure}[h]
	\begin{center}
		\vspace{-0.1in}
		\includegraphics[scale=0.48]{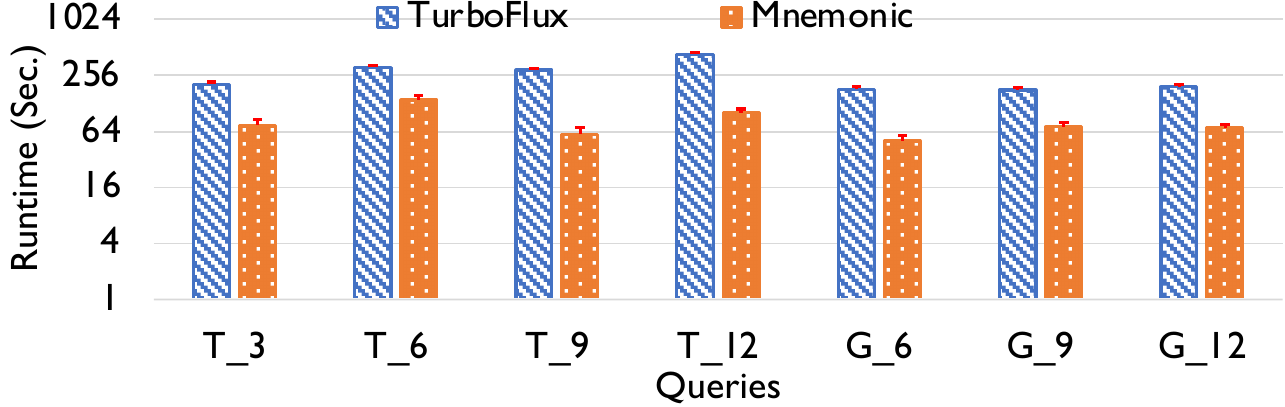}
		\vspace{-0.1in}
		\caption{Runtime of  TurboFlux and {\name}  on {\lsbench}}
		\vspace{-0.2in}
		\label{fig:vstf_lsbench}
	\end{center}
\end{figure}

\noindent \textbf{Insertion-deletion stream:} The runtime comparison for {\lsbench} is presented in Figure~\ref{fig:vstf_lsbench}. Edges are streamed in the batch of 16K. 
Both positive  and negative embeddings are listed. 
Negative embeddings are the matches of a query graph that should be removed after an edge deletion. 
{\name} is able to outperform TurboFlux by $3.27\times$ on average. The speedup is lower than for {\netflow} because it has fewer parallel edges and thus index structures become somewhat similar. Nonetheless, the advantage of {\name} is clear. Note that the runtime is more consistent here, likely because {\lsbench} is a random graph whereas {\netflow} follows power law. 

\begin{figure}[h]
	\begin{center}
		\vspace{-0.1in}
		\includegraphics[scale=0.48]{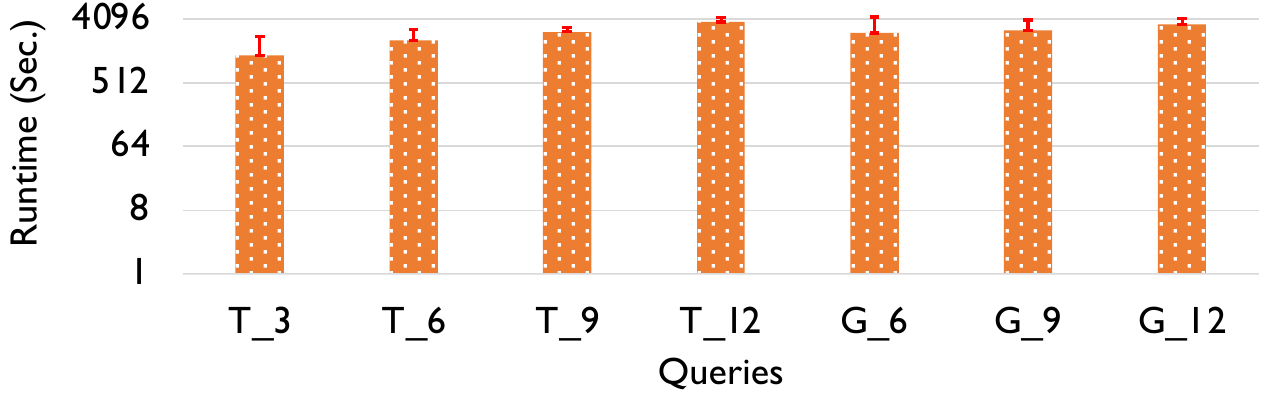}
		\vspace{-0.1in}
		\caption{Isomorphism on {\lanl} with sliding window }
		\vspace{-0.2in}
		\label{fig:vstf_lanl}
	\end{center}
\end{figure}

\noindent \textbf{Sliding-window stream:} 
 We only present the runtime of {\name} here as none of the existing systems were able to work out of the box under this scenario. The window and stride sizes are set to 24 hours and 10 minutes respectively.  
 The edges are repeatedly dropped from the tail of windows. The runtime for the isomorphic search of different query sizes is reported in Figure~\ref{fig:vstf_lanl}.  The runtime increases almost linearly to the size of queries. For three days of {\lanl} data, even the slowest queries are completed within two hours since search space is confined within the window. 

\begin{figure}[h]
	\begin{center}
		\vspace{-0.1in}
		\includegraphics[scale=0.50]{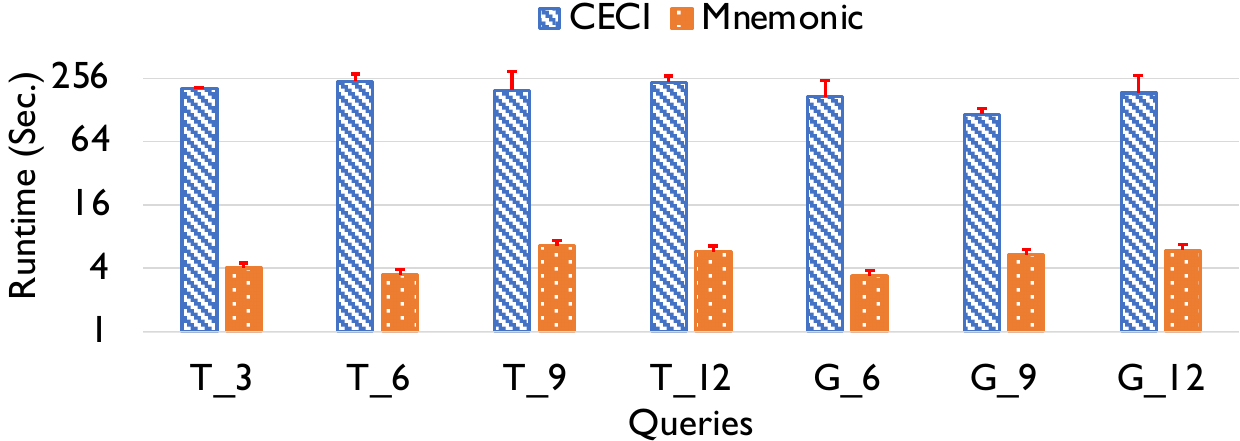}
		\vspace{-0.1in}
		\caption{Comparison of runtime per snapshot against CECI}
		\vspace{-0.2in}
		\label{fig:vsceci}
	\end{center}
\end{figure}

\noindent\textbf{Static Graph Solution:}
Figure~\ref{fig:vsceci} compares the average runtime per snapshot between {\name} and CECI~\cite{bhattarai2019ceci} on {\lanl}. While, we perform incremental computation, CECI performs subgraph matching from scratch on each snapshot. Using a 24-hour window and a 15-minute stride, 96 snapshots are generated starting at the beginning of day 2. Isomorphic search for all the queries is performed on every snapshot and the average runtime over all the snapshots is reported. As expected, {\name} easily outperforms CECI by $42\times$ per snapshot on average. However, on the first snapshot only CECI was slightly better( $\times1.08$) compared to {\name}. This shows that while dense index structures like CECI are better for enumeration(due to coalesced memory access), they are not efficient for streaming graphs.

\vspace{-2mm}
\subsection{Varying Search Granularity}
\label{sec:granularity}
\begin{figure}[h]
	\begin{center}
		\vspace{-0.1in}
		\includegraphics[scale=0.450]{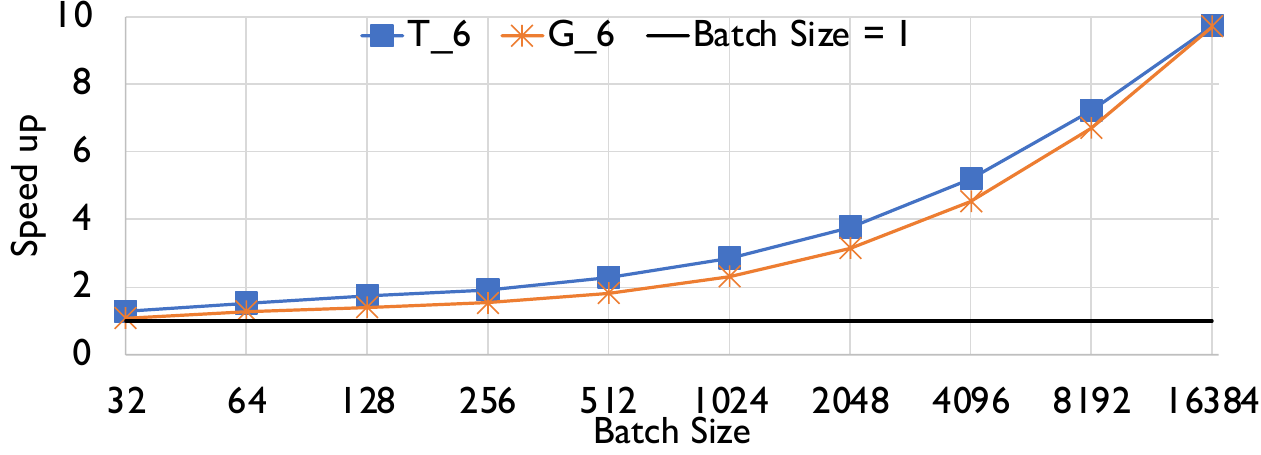}
		\vspace{-0.1in}
		\caption{Scalability over batch size.}
		\vspace{-0.2in}
		\label{fig:streamsize}
	\end{center}
\end{figure}

\vspace{-2mm}
Figure~\ref{fig:streamsize} shows  the average speedup for $G\_6$ and $T\_6$ as we increase the batch size in {\netflow} (stream size = 2M). Tree queries $T\_3$, $T\_6$, $T\_9$, and $T\_{12}$ achieve the maximum speedup of $4.2\times$, $9.7\times$, $10.6\times$, and $8.8\times$ respectively. Similarly, $G\_6$,  $G\_9$,  for $G\_12$ obtain the maximum speedup of $9.7\times$, $8.9\times$, and $10.2\times$ respectively. Note that we use the same number (1) of thread for all the batch sizes, to show that the benefit comes from shared traversal.

\begin{figure}[h]
	\begin{center}
		\vspace{-0.1in}
		\includegraphics[scale=0.50]{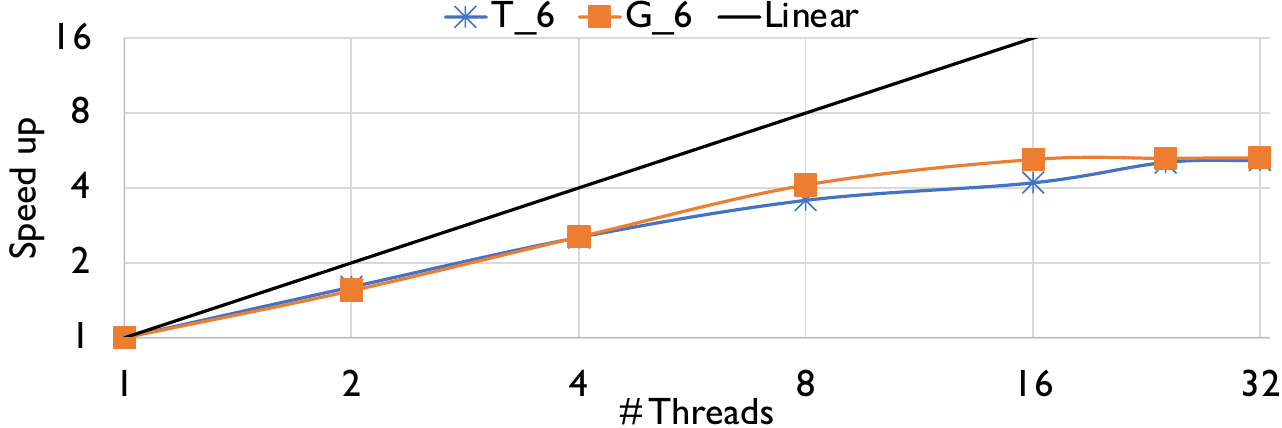}
		\vspace{-0.1in}
		\caption{Scalability over thread count.}
		\vspace{-0.2in}
		\label{fig:threadcount}
	\end{center}
\end{figure}

The frontier computation, top-down filtering, bottom-up filtering, and enumeration are all parallelized using \texttt{OpenMP}. Since {\idx} is indexed with \emph{edgeId},  both read and write are thread-safe, as two threads never process the same edge concurrently.  The speedup with increasing thread count for $T\_6$ and $G\_6$ is shown in Figure~\ref{fig:threadcount}, where the batch size  is fixed to 16K in {\netflow}(stream size = 2M). The average speedup among all queries for 24 threads is $5.22\times$.

\vspace{-3mm}
\subsection{Varying Embedding Types}
\label{sec:embeddingtype}
\begin{figure}[h]
	\begin{center}
		\vspace{-0.1in}
		\includegraphics[scale=0.55]{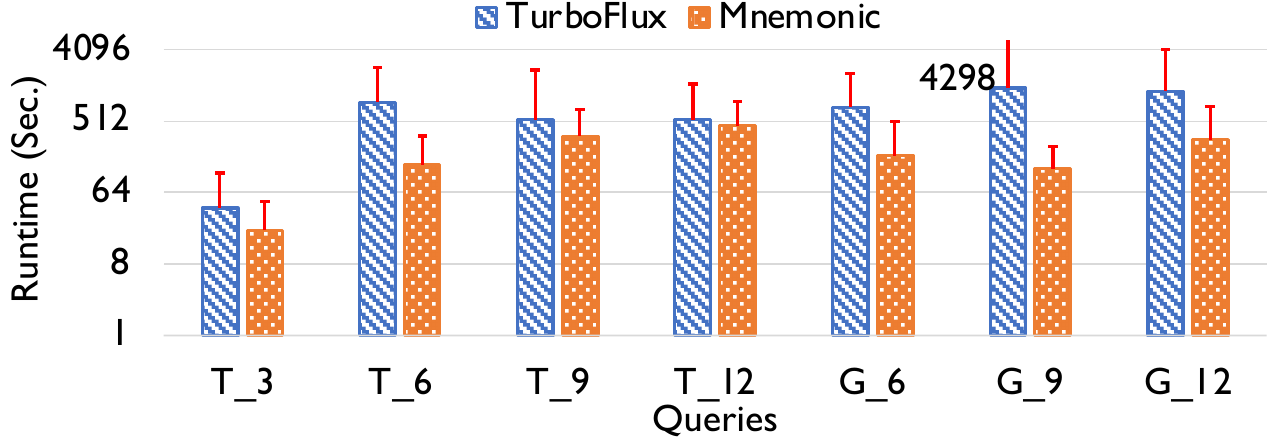}
		\vspace{-0.1in}
		\caption{Comparison of runtime between TurboFlux and {\name} for homomorphic enumeration}
		\vspace{-0.2in}
		\label{fig:vstf_netflow_homo}
	\end{center}
\end{figure}

\vspace{-1mm}
%
Figure~\ref{fig:vstf_netflow_homo} compares the runtime between {\name} and TurboFlux for \textbf{homomorphic enumeration} on the {\netflow} with 2M insertions. Since the it does not have to perform an injection check for each embedding, the process is faster compared to isomorphism and none of the queries timed out. {\name} is on average 4.2$\times$ faster than TurboFlux. 

\begin{figure}[h]
	\begin{center}
		\vspace{-0.1in}
		\includegraphics[scale=0.5]{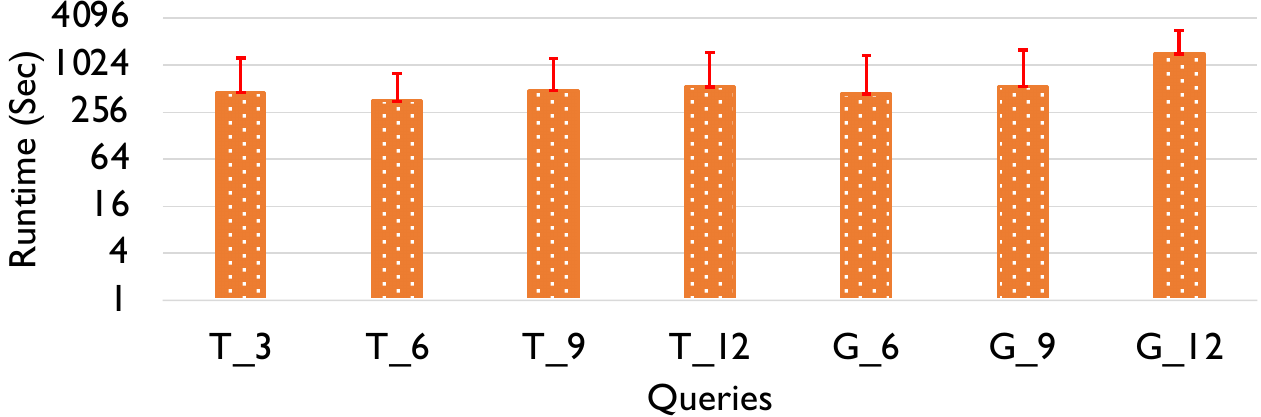}
		\vspace{-0.1in}
		\caption{Runtime for strong simulation on {\lanl} }
		\vspace{-0.2in}
		\label{fig:vstf_simulation}
	\end{center}
\end{figure}

Figure~\ref{fig:vstf_simulation} shows the runtime for \textbf{dual simulation} on the {\lanl} data graph. The window and stride sizes are set to 24 hours and 10 minutes respectively. The incremental simulation updates the {\idx} and computes a new binary relation from updated {\idx}. Due to relaxed constraints, the runtime for dual-simulation is much faster than isomorphism. Most of the queries are completed within 30 minutes. 

\begin{figure}[h]
	\begin{center}
		\vspace{-0.1in}
		\includegraphics[scale=0.5]{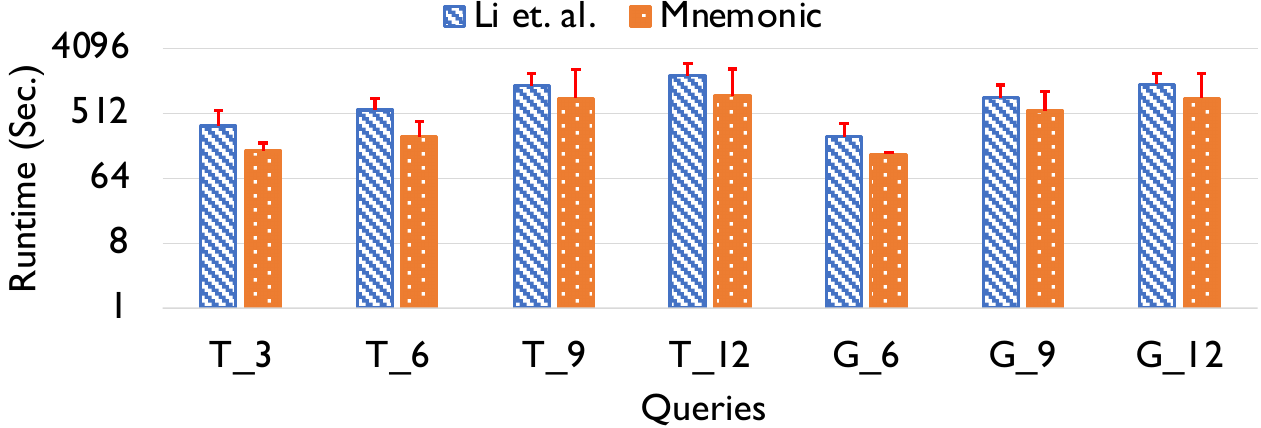}
	    \vspace{-0.1in}
		\caption{Runtime of {\name} and Li et.al.~\cite{li2019time} for time-constrained isomorphic search on \texttt{\lanl}}
		\vspace{-0.2in}
		\label{fig:vstf_temporal}
	\end{center}
\end{figure}

Lastly, we evaluate \textbf{time-constrained isomorphism}, where the edges in an embedding need to follow a specific temporal order that is encoded in the query graph via a timestamp. Figure~\ref{fig:vstf_temporal} compares the runtime of temporal-ordered isomorphism matching on {\name} against Li et al.~\cite{li2019time}. Both systems use the batch size of 16K on {\lanl}. {\name} on average is $1.8\times$ faster as the {\idx} is update friendly. In contrast, Li et al.~\cite{li2019time} needs to find the individual partially materialized results in their match-store tree and update them.


\vspace{-4mm}
\subsection{Memory Consumption}
\label{sec:memory}
\begin{figure}[h]
	\begin{center}
		\vspace{-0.1in}
		\includegraphics[scale=0.5]{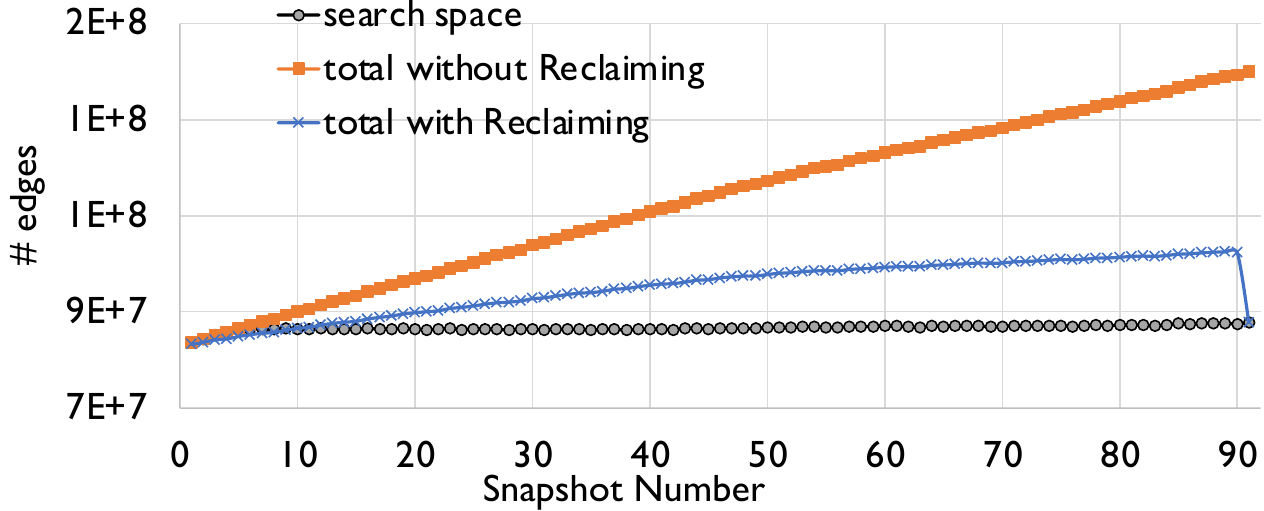}
		\vspace{-0.1in}
		\caption{Comparison of number of edge placeholders required to represent the active search context of the graph with and without memory reclaiming}
		\vspace{-0.2in}
		\label{fig:lanl_nf}
	\end{center}
\end{figure}

We measure the memory consumption trend over 90 snapshots starting from the beginning of day 2 in the {\lanl} data using the window size of 24 hours and stride of 10 minutes. Figure~\ref{fig:lanl_nf} shows the effect of memory reclaiming. Even though the number of events in the 24 hours window remains similar, the total number of edge entries (thus a placeholder in memory) increases rapidly. By reusing the memory of deleted edges, the growth in data size reduces from 67\% to 23\% over 90 snapshots. 
In addition, {\name} allows the periodic resets, where we can discard the cumulative index at any point in time (e.g., the $91^{st}$ snapshot) and rebuild the {\idx} starting from that snapshot.

For queries requiring a longer context, {\name} backs up older edges and {\idx} to the disks. We utilize the disk storage to run queries with a 3-day search window in {\lanl} where we only maintain events from last 24-hour in memory.  Table~\ref{tbl:diskvsmemory} shows the maximum space consumed in memory and disk as well as the overhead involved in maintaining {\idx} in disk and fetching candidates on {\enumerator}. While runtime suffers from higher candidate access latency, it allows to perform subgraph matching in large search window.

\begin{table}[h]
	\centering
	\caption{Storage-runtime trade off for disk based {\idx} }
	\vspace{-0.1in}
	\label{tbl:diskvsmemory}
	\begin{scriptsize}
	\begin{tabular}{|c|c|c|c|c|}
		\hline
		Query & \multicolumn{2}{c|}{Storage size required(GB)} & \multicolumn{2}{c|}{Overhead(\%)} \\ \hline
		& Memory              & Disk               & DEBI  mgmt.     & enumeration     \\ \hline
		T\_3  & 14.575              & 26.98              & 5.8\%           & 8.5\%           \\ \hline
		T\_6  & 26.1875             & 49.34              & 3.1\%           & 7.1\%           \\ \hline
		T\_9  & 39.8375             & 78.87              & 4.9\%           & 7.9\%           \\ \hline
		T\_12 & 52.7                & 102.13             & 8.2\%           & 9.2\%           \\ \hline
		G\_6  & 34.475              & 67.65              & 5.6\%           & 7.6\%           \\ \hline
		G\_9  & 53.375              & 106.87             & 6.7\%           & 7.8\%           \\ \hline
		G\_12 & 73.7                & 142.27             & 7.3\%           & 9.3\%           \\ \hline
	\end{tabular}
	\end{scriptsize}
\end{table}

The space requirement of {\idx} is $\mathcal{O}(|E|\times|V_q|)$ bits. Here $|E|$ is number of data edges, and $|V_q|$ is the number of query nodes. This is a sparse data structure, i.e., it needs to store the information between data and query edge pairs even if they do not match each other. For a data graph with 1 Billion edges and a query graph with 10 nodes, 10B bits i.e., 1.25 GB is needed for {\idx}. In a dense representation of graph like adjacency list or compressed sparse row, it takes $4\times10^9$ Bytes $\approxeq$ 4 GB for storing the graph, where 4 Bytes is used to represent a node id. With additional edge and node attributes, graph size is a much bigger space concern compared to {\idx}. Fortunately, with a vast pool of research focusing on external memory support for graphs, we leveraged one of those systems~\cite{zhu2019livegraph} to extend the {\name} beyond memory capability.

Dense candidate stores~\cite{bi2016efficient,bhattarai2019ceci}, which have similar worst space complexity, only store the data nodes that are potential matches of a given query node. Thus, the space advantage of dense stores is evident when the queries are very selective. They also provide a coalesced memory access while fetching the candidate matches of a query node in contrast to stride access pattern in {\idx}. This provides the dense storage approaches an advantage during enumeration. However, updating the dense storage requires searching through multiple key-value stores and each can take $\mathcal{O}(|V|)$. In {\idx}, we can add, remove and find the entries to be updated in constant time.




\section{Conclusion}
\label{sec:conclusion}
\vspace{-2mm}
{\name} presents a programmable subgraph matching system, that allows users to tailor the underlying components according to the nature of data and problem characteristics. Its utility on various types of data streams and different embeddings have been evaluated and compared against state of the art systems. We plan to open-source the system as soon as possible, as well as provide capability for distributed computation and querying multiple patterns concurrently as next additions. We strongly believe that it can democratize the process of implementing a subgraph matching system outside the small pool of researchers.

\bibliographystyle{IEEEtran}
\bibliography{Mnemonic}

\end{document}